\documentclass[%
 reprint,
 amsmath,amssymb,
 aps,
 preprintnumbers
]{revtex4-1}
\usepackage{graphicx}
\usepackage{dcolumn}
\usepackage{slashed}
\usepackage{bm}
\usepackage{tikz}
\usepackage{hyperref}
\usepackage{amsmath,amssymb}
\usetikzlibrary{decorations.pathmorphing}
\usetikzlibrary{decorations.markings}
\usepackage{color}
\usepackage{verbatim}
\usepackage{dsfont}

\begin{document}

\preprint{CERN-TH-2017-216 , IFT-UAM/CSIC-17-102}

\title{Renormalization group evolution of Higgs effective field theory}


\author{Rodrigo Alonso$^{a}$}
\author{Kirill Kanshin$^b$}
\author{Sara Saa$^c$}
\affiliation{$^a$Theoretical Physics Department, CERN, Geneva, Switzerland}
\affiliation{$^b $Dipartimento di Fisica e Astronomia `G. Galilei', Universit\`a di Padova INFN, Sezione di Padova, Via Marzolo 8, I-35131 Padua, Italy}
\affiliation{$^c $Departamento de F\'isica Te\'orica and Instituto de F\'isica Te\'orica, IFT-UAM/CSIC, Universidad Aut\'onoma de Madrid, Cantoblanco, 28049, Spain}



\begin{abstract}
The one-loop renormalization of the action for a set Dirac fermions and a set of scalars spanning an arbitrary manifold coupled via Yukawa-like and gauge interactions is presented.
The computation is performed with functional methods and in a geometric formalism that preserves at all stages the symmetries of the action.
The result is then applied to Higgs effective field theory to obtain the renormalization group evolution. 
In the Standard Model limit of this EFT the RGE equations collapse into a smaller linearly independent set; this allows to probe the dynamics of the scalar discovered at LHC via de-correlations in the running of couplings.
\end{abstract}


\maketitle

\section{Introduction}
Recent years have seen a resurgence in the study of effective field theories (EFT), a phenomenon motivated in large part by the goal of painting the Standard Model (SM)
as part of a bigger picture. The SM EFT as a case study has indeed pushed our understanding to new depths in ways worth reviewing;  {\it i)}
the reduction~\cite{Grzadkowski:2010es} of the original basis of operators~\cite{Buchmuller:1985jz}
raised the question of how to find the minimum complete set in the general case, solved via use of the conformal group~\cite{Henning:2015alf,Henning:2017fpj},
{\it ii)} the one-loop renormalization group evolution (RGE), now determined in full~\cite{Jenkins:2013wua,Jenkins:2013zja,Alonso:2013hga,Alonso:2014zka}, was shown to have an unexpected holomorphy structure~\cite{Alonso:2014rga} later accounted for with amplitude methods~\cite{Cheung:2015aba} (see also~\cite{Elias-Miro:2014eia}), {\it iii)} conventional Feynmann-diagram-based methods when used in large scale computations,  like the SM EFT RGE, proved cumbersome and inefficient;  in contrast  a  Covariant Derivative Expansion  (CDE) in conjunction with functional methods shows promise to trivialize one-loop computations~\cite{Gaillard:1985uh,Cheyette:1987qz,Henning:2014wua,Drozd:2015rsp,delAguila:2016zcb,Henning:2016lyp}. 

The SM EFT however is built on the assumption that the geometry of the Higgs doublet scalar-space is trivial, i.e. $\mathbb{R}^4$; allowing for more general geometries~\cite{Alonso:2016oah} leads to 
 what has come to be known as Higgs Effective Field Theory (HEFT) -- more prosaically the EFT for the longitudinal $W$ \& $Z$ modes~\cite{Appelquist:1980vg,Longhitano:1980iz,Longhitano:1980tm,Feruglio:1992wf} supplemented by Higgs singlet functions~\cite{Grinstein:2007iv}. 
The operator basis in this EFT has been laid out~\cite{Alonso:2012px,Buchalla:2013rka}, a process which itself triggered a revision of power counting~~\cite{Buchalla:2013eza,Jenkins:2013sda,Gavela:2016bzc}, and the one-loop computation added difficulties that this case entails~\cite{Gavela:2014uta}
have been addressed with a geometric description~\cite{Alonso:2015fsp}, and functional methods~\cite{Guo:2015isa}. The full one-loop RGE in HEFT has not been determined yet however; filling this gap is the aim of the present letter.

In this task we have put to use the aforementioned progress in the field; in particular the CDE and the geometric description of the HEFT are central to this work whereas we comment on selection rules from amplitude methods. 
It should be emphasized nevertheless that the results here presented not only apply to HEFT; they are valid for a 
a manifold of scalars and a set of fermions coupled via Yukawa interactions and subject to a gauged symmetry as made explicit in the next section.


\section{Gauge and Yukawa Theory for a Manifold of Scalars\label{TH}}
Consider a manifold parametrized by a set of $n_\phi$ scalar fields $\phi^i$ with metric $G_{ij}$, $n_\psi$ Weyl fermion fields  ($n_L$ left-handed (LH) and $n_R$ right-handed (RH)) $\psi$, a Yukawa-like coupling among
the two and a gauged symmetry group that acts on both.
The classical action reads:
\begin{align}\nonumber
S=\int d^4x\Bigg[ &\frac {G_{ij}}{2}  d^\mu\phi^i d_\mu\phi^j  -\frac14A_{\mu\nu}A^{\mu\nu}-V(\phi) \\
&+\bar\psi \left( i\slashed D -\mathcal M(\phi) \right)\psi\,\Big]\label{LOLag}\,,
\end{align}
where  $\mathcal M$ is decomposed in chiral projectors as $\mathcal M ={\mathcal M}_RP_R+{\mathcal M}_R^\dagger P_L$ with $\mathcal M_R$ a $n_L\times n_R$ matrix whose elements are functions of the scalar fields $\phi$.
The largest symmetry that the fermions can accommodate is that of the kinetic term, a unitary rotation in $n_\psi$ dimensions, $U(n_\psi)$;
as for the scalars, the kinetic term is invariant under isometry transformations, $\delta \phi^i= \theta^A t_A(\phi)$, given by the Killing vectors $t_A$ 
defined as:
\begin{align}
t_A^k\frac{\partial G_{ij}}{\partial\phi^k} +G_{kj}\frac{\partial t^k_A} {\partial \phi^i}+G_{ik}\frac{\partial t^k_A} {\partial \phi^j}&=0\,.
\end{align}
Considering Dirac masses, as we do,  automatically reduces the symmetry group in the fermion sector from $U(n_\psi)$ to $U(n_L)\times U(n_R)$.
The subgroup of the combination (isometries)$\times U(n_L)\times U(n_R) $  that is respected by the full action is given by the directions in Lie Algebra space ($\theta,\tilde\theta_L,\tilde\theta_R $), with $\delta\psi_{L(R)}=i\tilde\theta^A_{L(R)}T^{L(R)}_{A}\psi_{L(R)}$, $(T_A^{L(R})^\dagger =T_A^{L(R)}$, which satisfy
\begin{align}
&\theta^A t_A^i\frac{\partial V}{\partial \phi^{i}}=0\,,\\ \nonumber
&\theta^A t^i_A\frac{\partial \mathcal M_R}{\partial\phi^i}-iT^L_A\tilde \theta^A_L\mathcal M_R +i \mathcal M_R T^R_A\tilde\theta^A_R=0\,.
\end{align}

Part or the whole of this group is gauged, which we call $\mathcal G$, and the covariant derivatives~\cite{Alonso:2016oah} read:
\begin{align}\label{GCDdef}
d_\mu \phi^i=\partial_\mu \phi^i\!+\!A_\mu^B t^i_{B}(\phi),\quad \!\! D_\mu \psi=\left(\partial_\mu \!+\!i T_B^\psi A_\mu^B\right)\psi,
\end{align}
where the coupling constants are contained in the Killing vectors $t$ and  the hermitian generators $T^\psi$~\footnote{Alternatively the coupling constants are placed in the gauge bosons kinetic term.}, the latter contain chirality projectors $P_{L,R}$. The scalar sector is an arbitrary manifold and differential geometry will be used frequently; in particular our notation for the covariant derivative in field space is:
\begin{align}
\nabla_i S&=\frac{\partial S}{\partial \phi^i}\,, &
\nabla_i V^j&=\frac{\partial V^j}{\partial \phi^i}+\Gamma_{ik}^j V^k,
\end{align}
where $\Gamma^i_{jk}$ is the connection for the metric $G_{ij}$ and the space-time covariant derivative action on an upper index:
\begin{align}\label{DmuScalar}
\left({D}_\mu \eta \right)^i &= \left(\partial_\mu \eta^i + \Gamma^i_{kj} d_\mu \phi^k \eta^j\right)+A^B_\mu (\partial_j{t^i_B}) \eta^j .
\end{align}
In the equation above and in the following, we will use interchangeably $\partial/\partial \phi^i$ and $\partial_i$.
\subsection{one-loop Renormalization}
The one-loop ($\hbar$) correction in the path integral formalism takes the form of
a Gaussian integral centered around the EoM solution, $\Phi_0$, schematically:
\begin{align}
\int\!\! \mathcal D\Phi\, e^{iS[\Phi_0]+i\Phi^2\delta^2\!S[\Phi_0]/2+\mathcal O(\Phi^3)}\!=\!\frac{\mathcal Ne^{iS[\Phi_0]}}{\sqrt{\det(-\delta^2S[\Phi_0])}},
\end{align}
so that the correction to the action reads $i$Tr$[\log(-\delta^2S[\Phi_0])]/2$. The Coleman-Weinberg potential can be obtained for background fields $\Phi_0$ taken as constants;
the generalization to space-time dependent fields requires of a Covariant Derivative Expansion (CDE)~\cite{Gaillard:1985uh,Cheyette:1987qz} recently realised to its full potential~\cite{Henning:2014wua,Drozd:2015rsp,delAguila:2016zcb,Henning:2016lyp}.
The application of this program to our case has two subtleties: {\it i)} the scalar fields have a non-flat geometry and {\it ii)} there are cross-terms with different spin-species fields in the second variation of the action. Point {\it i} is addressed with a covariant formulation and in particular a covariant second variation $\nabla^2S$~\cite{Alonso:2015fsp}, as for the addressing of {\it ii},  the procedure is completing squares via field redefinitions which do not change the measure~\cite{Henning:2016lyp}. 
After implementing the above steps, the second variation of the action reads, in the Feynman-t'Hooft gauge:
\begin{align}
\nabla^2S&=\delta\bar  \psi\Pi_\psi\delta\psi-\frac12\delta A_\mu
\hat\Pi_V^{\mu\nu}
\delta A_\nu
 \\ \nonumber
&-\frac12\delta \phi \left\{ \Pi_{\phi}+2 \bar\psi \nabla \mathcal M \Pi_\psi^{-1}\nabla \mathcal M\psi -l_\mu\left(\hat \Pi_{V}^{\mu\nu}\right)^{-1}\!r_\nu \right\}\delta\phi,
\end{align}
where we have omitted scalar and gauge summed indices, made explicit in the definitions
\begin{align}\label{PIA}
&\Pi_\psi= i\slashed D-\mathcal M ,\\ \nonumber
&\left[\Pi_\phi\right]_{ij}=[D^\mu D_\mu]_{ij} +R_{ikjl}d_\mu\phi^k d^\mu\phi^l+\nabla_j\nabla_i( V+\bar\psi \mathcal M\psi ),\\ \nonumber
&[\Pi_V^{\mu\nu}]_{AB}\!=\delta_{AB}(-g^{\mu\nu}) D^\rho D_\rho -2f_{AB}^C A_C^{\mu\nu}-g^{\mu\nu}t_A t_B, \\ \nonumber
&[\hat \Pi_V^{\mu\nu}]_{AB}\!=[\Pi_V^{\mu\nu}]_{AB}+\bar\psi \gamma^\mu T_A \Pi_\psi^{-1} \gamma^\nu T_B\psi+h.c.,\\ \nonumber
&l^{\mu\,i}_A=-t_A^i  D^\mu +2 d^\mu \phi   \nabla^i  t_A-\bar\psi \nabla^{i}\mathcal M  \Pi_\psi^{-1}  \gamma_\mu T_ A \psi+h.c.,\\ \nonumber
&r^{\nu\,i}_B=D^\nu t_B^i +2d^\nu \phi  \nabla^{i} t_B-\bar \psi \gamma^\nu T_ B\Pi_\psi^{-1}\nabla^{i}\mathcal M  \psi+h.c.,
\end{align}
where $[T^\psi_B,T^\psi_C]=if^{A}_{\,\,BC}T_A^\psi$ are the structure constants, $D_\mu$ as in eqs.~(\ref{GCDdef}),(\ref{DmuScalar}) -- in particular in eq.~(\ref{PIA}) it acts on the adjoint representation with  $(T_A^\mathcal G)^B_{\,\,C}=if^B_{\,\,AC}$\,-- 
and $R_{ijkl}$ is the Riemann tensor in scalar space.
The final result then reads in dimensional regularization, $\int d^4x \mathcal P/(16\pi^2(4-{\rm d}))$ with: 
\begin{widetext}
\begin{align}\label{1LGeneral}
\mathcal P=&\bar\psi T \left(2i\slashed D-8\mathcal M\right) T\psi+\frac{11}{12}\mathbb C_{\mathcal G}  A_{\mu\nu}A^{\mu\nu}+ \frac{1}{2}\mbox{Tr}\big[\left(R d\phi^2\!+\nabla^2(V+\bar \psi \mathcal M \psi)-t\cdot t\right)^2\! +\frac{1}{6}[D,D]^2\!+2(t\cdot t)^2\big]   \\ \nonumber
&+\bar\psi \nabla \mathcal M \left(i\slashed D+2\mathcal M^\dagger \right)\nabla \mathcal M\psi-\left(i2\bar\psi t \nabla\mathcal M   T \psi +h.c.\right)
-2d_\mu\phi \nabla_i t d^\mu\phi \nabla^i t -\frac12\mbox{Tr}\left\{\left(\mathcal M^\dagger \mathcal M -i(\slashed D\mathcal M)\right)^2\!-\frac{1}{6}[D,D]^2 \right\},
\end{align}
\end{widetext}
where $\mathbb C_\mathcal G$ is the Casimir of the adjoint representation  $\sum T_{A}^\mathcal G T_A^\mathcal G=\mathbb C_\mathcal G \mathds 1$ and traces are on scalar and Dirac space respectively.
This is the set of counter-terms required for the action of eq.~(\ref{LOLag}) at the one-loop level and the central result of this work. 
A number of well known results can be recovered from here, e.g. the SM which will be discussed in sec.~\ref{HEFT} as a limit of HEFT.

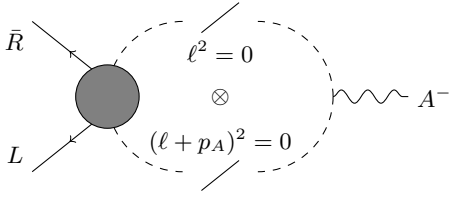
\begin{figure}
\begin{tikzpicture}
\draw [style=dashed] (1,1) arc (90:180:1);
\draw [style=dashed] (0,0) arc (180:270:1);
\draw [style={postaction={decorate}, decoration={markings,mark=at position .5 with{\arrow{>}}}}] (0,0.2)--(-1,1) node[anchor=north east] {$\bar R$};
\draw [style={postaction={decorate}, decoration={markings,mark=at position .5 with{\arrow{>}}}}] (0,-0.2)--(-1,-1) node[anchor=south east] {$L$};
\filldraw[gray,draw=black] (0,0) circle (12pt);
\draw (1.5,0) node {$\otimes$};
\draw [style=dashed] (3,0) arc (0:90:1);
\draw [style=dashed] (3,0) arc (0:-90:1);
\draw [style={decorate}, decoration={snake}] (3,0)--(4,0);
\draw (1.5,.6) node {$\ell^2=0$};
\draw (1.5,-.6) node {$(\ell+p_A)^2=0$};
\draw (1.25,-1.25)--(1.75,-0.85);
\draw (1.25, 0.85)--(1.75,1.25);
\draw (4.35,0) node {$A^-$};
\end{tikzpicture}
\caption{Cut on the diagram built with on-shell amplitudes that would produce a dipole term, the arrows on the fermion lines indicate helicity.\label{HelPsiA}}
\end{figure}

Despite it being a lengthy expression one can readily see in eq.~(\ref{1LGeneral}) that dipole or three field strengths terms are absent. It is simple to realise~\cite{Buchalla:2013rka} that the 1-loop diagrams with three external gauge bosons do not give a divergent contribution to $F^3_{\mu\nu}$ operators given the operator momentum dependence. For the dipole terms, the action in eq.~(\ref{LOLag}) produces a contact $\bar\psi_L\psi_R\phi\phi$ vertex whose insertion in the diagram of fig.~\ref{HelPsiA} could yield a divergent contribution to the dipole. However a look at the Dirac structure of the fermion bilinear in the diagram reveals no piece proportional to $\sigma_{\mu\nu}$. It is worth looking at how the absence of dipole terms is deduced with amplitude methods for contrast.

First we note that dipole operators generate vertices with the sum of helicities -- with all particles taken e.g. outgoing --
of $\pm 2$; no term with such property appears in eq.~(\ref{1LGeneral}).
To find out if such a dipole term is generated by RGE one can look into the cuts of diagrams built out of on-shell amplitudes~\cite{Bern:1994cg,Bern:1994zx}, diagrammatically shown in fig.~\ref{HelPsiA}.
In the case in which $\mathcal M$ is linear in $\phi$ the amplitude $\mathcal A(L\bar R\phi\phi )$ cancels but not in the general case. 
We therefore pursue further the computation;  the on-shell amplitudes momentum dependence is, in the massless limit, $\mathcal A(L\bar R\phi\phi )\sim\langle L\,R\rangle$, $\mathcal A(A^-_\mu\phi^i\phi^j )\sim\langle A\,i\rangle\langle j\,A\rangle/\langle i\,j\rangle \partial_{i}t_{j}^A$, where $|k\rangle$ is the Weyl spinor for a helicity $-1/2$ massless particle of momentum $k$ and the scalar product $\langle k\, k^\prime\rangle$ is antisymmetric, and finally imposing in the internal scalars the on-shell conditions of fig.~\ref{HelPsiA} the result is:
\begin{align}
\sum_{ij}\left\langle L \partial_i \partial_j \mathcal M R \right\rangle (\partial_it^A_j- \partial_j t^A_i)\frac{\left\langle AL\right\rangle\left\langle AR\right\rangle}{\left\langle L R\right\rangle},
\end{align}
so that the sum in scalar indices cancels and no dipole term is produced by RGE.
\section{Higgs EFT\label{HEFT}}
The previous result can be applied to Higgs EFT. The gauge group is $\mathcal G=SU(3)_c\times SU(2)_L\times U(1)_Y$ and there are six fermion
representations, which we group in \begin{align} 
\psi_L&=(q_L,\ell_L)^T, & \psi_R&=(u_R, d_R,\nu_R,e_R)^T,\\ \nonumber
Q_{\psi_L}\!&=\text{Diag}\left(\frac16,-\frac12\right),
&Q_{\psi_R}\!&=\text{Diag}\left(\frac23,-\frac13,0,-1\right),
\end{align} where RH neutrinos are assumed Dirac, $T_Y^\psi=g_{\scriptscriptstyle Y}Q_{\psi}$ in our gauge derivative of eq.~(\ref{GCDdef}) and the non-abelian part connects with usual conventions as Tr$(T_AT_B)={g^2_{(s)}}\delta_{AB}/2$ for weak isospin (color).
The scalar sector has 4 fields which we arrange in
$\phi=(\varphi^a\,,\,h)$; where $h$ is the Higgs singlet and $\varphi^a$, $a=1,2,3$, the Goldstones to be eaten by the $W$ and $Z$. 
The metric reads:
\begin{align}
&G_{ij}=\mbox{Diag}(F(h)^2 g_{ab}(\varphi)\,,\,1)\,,
\end{align}
with $F$ a generic function of the singlet $h$. To specify the metric in Goldstone space, $g_{ab}$, which in turn determines $M_Z/M_W$, we assume custodial invariance which is equivalent to
taking the Goldstone bosons manifold to be $S^3$~\cite{Alonso:2016oah} which admits an $SO(4)=SU(2)\times SU(2)$ symmetry. To write the metric and Goldstone couplings
it is useful to introduce:
\begin{align}\nonumber
g_{ab}&=\partial_a u(\varphi) \cdot \partial_b u(\varphi)& t^i_A&=(g^{ab}\partial_b u\cdot \hat T_A \cdot u\,,\,0)\\
U(\varphi)&=\hat \sigma \cdot u(\varphi)
& \hat \sigma &\equiv\left\{i\sigma^i\,,1\right\}\\ \nonumber
u(\varphi)& \cdot u(\varphi)=1 &U&U^\dagger=U^\dagger U=1
\end{align}
where $u(\varphi)$ is an auxiliary unit vector in $\mathbb{R}^4$ parametrized by $\varphi$ with $u^j(0)=\delta^j_{4}$
and $\hat T_A$ are antisymmetric real matrices which, given the usual convention, satisfy Tr$(\hat T_A\hat T_B)=-g_{({\scriptscriptstyle Y})}^2\delta_{AB}$ for weak isospin (hypercharge).
With these definitions, the Lagrangian reads
\begin{align}\label{LagHLO}
\mathcal L=& \frac12 \partial_\mu h \partial^\mu h+\frac12F(h)^2d\varphi^2 - \frac14A_{\mu\nu}A^{\mu\nu}-V(h) \\ \nonumber
&+\bar\psi \left( i\slashed D -\mathcal U(\varphi) \, \mathcal  Y(h)P_R- \mathcal  Y^\dagger(h) \,\mathcal U^\dagger(\varphi)P_L \right)\psi \,,
\end{align} 
where $\mathcal Y=$Diag$(\mathcal{Y}_u,\mathcal{Y}_d,\mathcal{Y}_\nu,\mathcal{Y}_e)$ as imposed by hyper-charge with each $\mathcal{Y}_I$ a $3\times 3$ matrix in flavor space and
$\mathcal U=\mathds 1\otimes U\!=$\,Diag$(U,U)$.
It should be underlined that $h$ is the excitation around the EW vacuum and therefore  $V^\prime(0)=0$, $F(0)\equiv v=2M_W/g=246$ GeV.
For illustration and later use we define here the SM as:
\begin{align}\nonumber
&F_{\rm SM}=h+\left\langle h\right\rangle,  \quad \mathcal Y_{\rm SM}=\frac{h+\left\langle h\right\rangle}{\sqrt{2}} Y,\quad \left\langle h\right\rangle^2=\frac{2m_h^2}{\lambda},\\ 
&V_{\rm SM}=-\frac{m_h^2}{2}\left(h+\left\langle h\right\rangle \right)^2+\frac{\lambda}{8}\left(h+\left\langle h\right\rangle \right)^4 ,\label{SMLim}
\end{align}
where we see that everything can be rewritten in terms of the scalar field around the $O(4)$-symmetric point $\bar h=h+\left\langle h\right\rangle$ and one has $\left\langle h\right\rangle=v$.

\subsection{Renormalization of the leading Lagrangian}

The original Lagrangian in eq.~(\ref{LagHLO}) receives corrections in all terms at one-loop. These corrections are Higgs-singlet field dependent,
e.g. $K(h)(\partial_\mu h)^2$, except for the gauge kinetic term. It is then required, in order to revert the Lagrangian
to its original form, to have $h$-dependent renormalization factors $Z(h)$. Explicitly:
\begin{align} \nonumber
\psi_R\to\psi_R-\frac{d_\epsilon}{32\pi^2}&\left(\mathbb C _{\psi_R}^\mathcal G+\frac{1}{2}\left(3\frac{\mathcal{Y}^\dagger \mathcal{Y}}{F^2}+\mathcal{Y}^{\prime\dagger}\mathcal{Y}^\prime \right)\right)\psi_R
\end{align}
\begin{align}  \nonumber
\psi_L\to\psi_L-\frac{d_\epsilon}{32\pi^2}&\Bigg(\mathbb C_{\psi_L}^\mathcal G +\frac{\mathcal U}{2}\left(\mathcal{Y}^\prime \mathcal{Y}^{\prime\dagger}-\frac{\mathcal{Y}\mathcal{Y}^\dagger}{F^2}\right) \mathcal U^\dagger\\&
+\frac{\mathcal{Y}\mathcal{Y}^\dagger+\mathcal{\tilde Y}\tilde{\mathcal{Y}}^\dagger}{F^2}\Bigg)\psi_L \label{hRen}
\end{align}
\begin{align}\nonumber
h\to h-\frac{d_\epsilon}{32\pi^2}& \int dh \Bigg[\frac{(g^\prime)^2+3g^2}{4}\left(F^{\prime\prime}F-2(F^{\prime})^2\right)\\ \nonumber
&+2\mbox{Tr}( \mathcal{Y}^\prime \mathcal{Y}^{\dagger\prime})-3\frac{V^\prime F^{\prime\prime}F^\prime}{F^2}\Bigg] 
\end{align}
where $d_\epsilon=2/(4-{\rm d})+\log\mu^2$,  $\mathbb C_\mathcal R^{\mathcal G}$ is the Casimir of the representation $\mathcal R$, $\sum_\mathcal G T_A^\mathcal R  T_A^\mathcal R =\mathbb C_\mathcal R^{\mathcal G}\mathds 1$ , e.g. $\mathds C_{q_L}=(1/6)^2g_{\scriptscriptstyle Y}^2+3 g^2/4+4g_s^2/3$
and $\tilde{ \mathcal Y}=( \mathds 1\otimes \varepsilon) \mathcal Y=$Diag$(\varepsilon$ Diag$(\mathcal{Y}_u\,,\mathcal{Y}_d), \varepsilon$ Diag$(\mathcal{Y}_\nu\,,\mathcal{Y}_e)$) with $\varepsilon$ the antisymmetric 2-tensor.

Gauge bosons have the SM renormalization, which can be read off from eq.~(\ref{1LGeneral}); we therefore do not make explicit the RGE of $g_{\scriptscriptstyle Y}, g$ and $g_s$ which can be found in e.g.~\cite{Arason:1991ic}.

After these redefinitions there are only 3 terms left in the Lagrangian that differ from the classical action; Yukawa couplings,
the potential and the kinetic term for the Goldstone bosons $\varphi$. The corrections received are again functions of the Higgs-singlet
so that here we present the RGE as {\it  differential equations in the renormalization scale $\mu$ of functions of the Higgs singlet} -- for phenomenological applications is best to regard the following as RGE for the $h$-Taylor expansion coefficients:
\begin{widetext}
\begin{align}\label{RGEF}
\mu\frac{d F(h,\mu)^2}{d \mu}-\frac{FF^\prime}{16\pi^2}\int\left[\frac{g_{\scriptscriptstyle Y}^2+3g^2}{2}\left(F^{\prime\prime}F-2(F^{\prime})^2\right)+4\mbox{Tr}(\mathcal{Y}^\prime \mathcal{Y}^{\prime\dagger})-6\frac{V^\prime F^{\prime\prime}F^\prime}{F^2}\right]dh&\\ \nonumber
+\frac{2}{16\pi^2}\left[2\mbox{Tr}(\mathcal{Y} \,\mathcal{Y}^\dagger)-\frac{g_{\scriptscriptstyle Y}^{ 2}+3g^2}{2}F^2+F^2(F^{\prime2}-1)\left(\frac{g_{\scriptscriptstyle Y}^{2}}{4}-2\frac{V^\prime F^\prime}{F^3}\right)-FF^{\prime\prime}V^{\prime\prime}\right]&=0\,,
\end{align}
\begin{align}\label{RGEV}
&\mu\frac{d V(h,\mu)}{d\mu}-\frac{V^\prime}{32\pi^2} \int \left[\frac{g_{\scriptscriptstyle Y}^2+3g^2}{2}\left(F^{\prime\prime}F-2(F^{\prime})^2\right)+4\mbox{Tr}( \mathcal{Y}^\prime \mathcal{Y}^{\prime\dagger})-6\frac{V^\prime F^{\prime\prime}F^\prime}{F^2}\right]dh\\ \nonumber
&-\frac{1}{16\pi^2}\Bigg[
\frac12(V^{\prime\prime})^2+\frac32\frac{(V^\prime F^\prime)^2}{F^2}-2\mbox{Tr}\left(\mathcal{Y}^\dagger \mathcal{Y}\right)^2-\frac{g_{\scriptscriptstyle Y}^{2}+3g^2}{4}FF^\prime V^\prime+\frac32F^4 \frac{g_{\scriptscriptstyle Y}^{4}
+2g_{\scriptscriptstyle Y}^{2}g^2+3g^4}{16}\Bigg]=0\,,
\end{align}
\begin{align}\label{RGEY}
-&\mu\frac{d \mathcal{Y}(h,\mu)}{d \mu}+\frac1{16\pi^2}\Bigg[g_{\scriptscriptstyle Y}^{2}\left(Q_{\psi_L}^2\mathcal{Y}+\mathcal{Y}Q_{\psi_R}^2-8Q_{\psi_L}\mathcal{Y}Q_{\psi_R}-\frac {\mathcal{Y}}{4}+\mathcal{Y}^\prime\left(\int \frac{F^{\prime\prime}F-2F^{\prime2}}{4}dh-\frac{F^\prime F}{4}\right)\right) \\ \nonumber
&-6\mathcal{Y}\mathbb C_{\psi}^{SU(3)_c}+3g^2\mathcal{Y}^\prime\left(\int \frac{F^{\prime\prime}F-2F^{\prime2}}{4}dh-\frac{F^\prime F}{4}\right)+\frac12\left(\mathcal{Y}^{\prime}\mathcal{Y}^{\prime\dagger}\mathcal{Y}+\mathcal{Y}\mathcal{Y}^{\prime\dagger}\mathcal{Y}^\prime \right)+2\mathcal{Y}^{\prime}\mathcal{Y}^\dagger \mathcal{Y}^{\prime} \\ \nonumber&
-3\frac{\tilde {\mathcal{Y}} \tilde {\mathcal{Y}}^\dagger}{F^2} {\mathcal{Y}}+{\mathcal{Y}}^\prime \int \left(2 \mbox{Tr}\left({\mathcal{Y}}^\prime {\mathcal{Y}}^{\prime\dagger}\right)-3 \frac{V^\prime F^\prime F^{\prime\prime}}{F^2}\right)dh+V^{\prime\prime} {\mathcal{Y}}^{\prime\prime}+3\frac{V^\prime F^\prime}{F^3}({\mathcal{Y}}^\prime F^\prime F-{\mathcal{Y}})\Bigg]=0\,.
\end{align}
\end{widetext}
As we remarked at the beginning of this section, we are expanding around the true vacuum so that $V^\prime(0)=0$; {\it to maintain this condition} at the Loop level we have to
make sure the RGE for the the linear term in $h$ cancels in eq.~(\ref{RGEV}). For this {\it we use the renormalization of the Higgs field} in eq.~(\ref{hRen}) choosing the limits of 
integration so as to cancel the RGE. It is a check on the procedure and worth examining how this works in the SM limit of eq.~(\ref{SMLim}). In this case one can define $\bar h=h+\left\langle h\right\rangle$ and the vev disappears from all RGE and in particular eq.~(\ref{RGEF}) for $F^2$ becomes a trivial $0=0$. Then after solving the remaining equations for the potential and Yukawas the defintion of $\left\langle h\right\rangle$ has to be revised which is simply taking $\lambda$ and $m_h$ in the expression for the vev, $\left\langle h\right\rangle^2=2m_h^2/\lambda$, to be running constants.
 If one instead follows the procedure above expanding around the true vacuum and imposing tadpole cancellation, the RGE for $F^2$ is not trivial and gives us the running of $\left\langle h\right\rangle$, which is precisely the running one would get for $2m_h^2/\lambda$ via the previous procedure.

From the perspective of the RGE of eqs.~(\ref{RGEF})-(\ref{RGEY}) the renormalizability of the SM follows from a number of cancellations and is a very special case. To illustrate this point let us look at the coupling $hW W$; in the SM this is $gM_W$ and its running is given by that of $g$ and $M_W$, however in HEFT it runs independently; if we write $\left(v^2+2 a v h \right) g^2W_\mu^+W_-^\mu/4\subset\mathcal L$ and consider a top Yukawa interaction as $\mathcal Y_t=(y_t^0 v+y_t^1 h)/\sqrt2$, $a$ runs as:
\begin{align}
\mu\frac{\partial a}{\partial \mu}=&\frac{N_c y_t^1}{8\pi^2}\left(y_t^1-y_t^0\right),
\end{align}
where given the poor current experimental precision on the coupling $h\bar tt$ one could have a 10\% effect for one order of magnitude scaling.
This illustrates how looking at the de-correlation of running of couplings can be used to probe scalar dynamics; de-correlations are indeed an experimental feature of HEFT~\cite{Brivio:2013pma}.

\subsection{Renormalization of the Sub-leading Lagrangian}

For consistency in our loop expansion, the coefficients of the NLO operators
have a $1/16\pi^2$ factor~\cite{Manohar:1983md}
and the RGE is simpler than in the case of the LO Lagrangian
since field normalization is a two-loop effect. Let us write
\begin{align}
\mathcal L_{NLO}=\frac{C_\alpha}{16\pi^2} \mathcal O_\alpha+\left[\frac{C_\alpha}{16\pi^2} \bar{\mathcal  O}_\alpha+h.c.\right],
\end{align}
where $\alpha$ indexes operators, themselves split into hermitian ($\mathcal O$) and non-hermitian ($\bar{\mathcal O}$).
There is one operator that is naively LO but is naturally suppressed
since it violates custodial symmetry, $\mathcal O_{\slashed C}\equiv\left(d_\mu \varphi t_Y\right)^2$.
Even if the new physics generating the HEFT at the scale $\Lambda$ respects custodial symmetry the RGE induced by hypercharge does not, and the operator is generated by the flow
\begin{align}
\mu\frac{d C_{\slashed C}}{d \mu}-3F^2(F^{\prime2}-1)=0\,,
\end{align}
where $g_{\scriptscriptstyle Y}$ does not show explicitly due to the defintion of $\mathcal O_{\slashed C}$. This operator contributes to the $T$ parameter~\cite{Peskin:1990zt} as
\begin{align}
\alpha\,\delta T=-\frac{3g_{\scriptscriptstyle Y}^2}{32\pi^2} (F^{\prime2}(0)-1) \log\left(\frac{v}{\Lambda}\right),
\end{align}
which has to be below the permile level making it natural to assume that $\mathcal O_{\slashed C}$ is NLO.

The remaining sub-leading Lagrangian needed for renormalization can be given in terms of 20 operators 
which have indices in flavor and fermion-species~\footnote{The fermion-species index runs through $I=u,d,\nu,e$, which means that the number of operators as customary defined~\cite{Grzadkowski:2010es} is larger than 20}. Non-abelian gauge indices are omitted
and contracted to build invariants so e.g. the combination $\mathcal U ^\dagger \psi_L$, $(\bar\psi_L \mathcal U)$
has the $SU(2)_L$ indices summed over and has a `RH' index and a flavor index $(\mathcal U ^\dagger \psi_L)_I^{\alpha}$,
 $I=u,d,\nu,e$ and $\alpha=1,2,3$. On the other hand $U(1)_Y$ forces some of the fermion species indices to be the same, e.g. $(\bar\psi_L \mathcal U)^I\psi_R^J$
only is $U(1)_Y$-invariant for $I=J$. The list reads -- the indices $A,B$ run through $SU(2)_L\times U(1)_Y$:
\begin{align} \nonumber
\mathcal O_{1}\!& =\!(\partial_\mu h)^4/F^4, & \mathcal O_{2}\!& =\! (d_\mu\varphi\cdot d_\nu\varphi)^2,\\ \nonumber
 \mathcal O_{3}\!& = \left((d_\mu\varphi)^2\right)^2,& \mathcal O_{4}\!&=(\partial_\nu h)^2 d_\mu\varphi^2/F^2,\\\nonumber
\mathcal O_{5}\!&= (d_\mu\varphi\partial^\mu h)^2/F^2, & \mathcal O_{6}\!& =d_\mu \varphi^a  d_\nu \varphi_b \nabla_a t^b_B A_{\mu\nu}^B,   \\  \nonumber
\mathcal O_{7}\!& =\partial^\mu h d^\nu \varphi \cdot t_B  A_{\mu\nu}^B/F, \! & \mathcal O_{8}&= t_B t_C A_{\mu\nu}^B A^{C,\mu\nu},\\\nonumber
\bar {\mathcal  O}_{9}\!&= (\partial_\mu h/F)^2( \bar \psi_L \mathcal U)^I  \psi_R^I, &\bar{ \mathcal O}_{10}^I\!&=  (d_\mu \varphi)^2 (\bar \psi_L \mathcal U)^I \psi_R^I,\\
\bar  {\mathcal O}_{11}^I\!&= \partial^\mu hd_\mu\varphi^a \partial_a(\bar\psi \,\mathcal U)^I\psi_R^I/F,& &
   \end{align}
\begin{align} 
\mathcal O^{*,\,I}_{12}&=\partial_\mu h\bar\psi^I \gamma_\mu\psi_R^I/F, \\ \nonumber
\mathcal O^{*,\,I}_{13}&=\partial_\mu h(\bar\psi \,\mathcal U)^I\!\gamma^\mu(\mathcal U^\dagger \psi_L)^I\!/F,\\ \nonumber
 \mathcal O_{14}^{IJ}&=id_\mu \varphi^a\bar\psi^I\gamma^\mu\left(\mathcal U^\dagger \partial_a \mathcal U \right)^{IJ}\psi_R^J ,\\ \nonumber
 \mathcal O_{15}^{I}&=id_\mu\varphi^a (\bar \psi \,\mathcal U)^I\gamma^\mu \overleftrightarrow\partial_a (\mathcal U^\dagger \psi_L)^I/2 ,\\ \nonumber
\mathcal O^{*,\,I}_{16}&= d_\mu\varphi^a \partial_a (\bar \psi\, \mathcal U)^I\gamma^\mu  (\mathcal U^\dagger\psi_L)^I/2 ,
\end{align}
\begin{align}\nonumber
\mathcal O_{17}^{IJ}&\!=\!(\bar\psi_L \mathcal U)^I \psi_R^I\,\bar\psi_R^J( \mathcal U^\dagger\psi_L)^J,\!  &\bar{ \mathcal O}_{18}^{IJ}&\!= \!(\bar\psi_L \mathcal U)^I \psi_R^I (\bar\psi_L \mathcal U)^J \psi_R^J,\\ \nonumber
\mathcal O_{19}^{IJ}&\!=\!\bar\psi_R^I  [\psi_L \bar\psi_L\otimes \mathds 1]^{IJ}  \psi_R^J,\!  &
\bar{\mathcal O}_{20}^{IJ}&\!=\![\bar\psi_L  (\psi_R^I) \bar\psi_L\otimes \varepsilon ]^{IJ} \psi_R^J,\\ \nonumber
\end{align}
where in the tensor products $SU(2)_L$ indices are contracted; they read, explicitly, $(\psi_L \bar\psi_L\otimes \mathds 1)^{IJ}=\{(q\bar q_L\mathds 1,q\bar\ell_L\mathds 1),(\ell\bar q_L\mathds 1,\ell\bar\ell_L\mathds 1)\}$ whereas in $\bar {\mathcal O}_{20}$
we have $[ \bar\psi_L (\,)\bar\psi_L \varepsilon]^{IJ}=\{(\bar q_L(\,)\bar q_L\varepsilon,\bar q_L(\,)\bar\ell_L\varepsilon),(\bar\ell(\,)\bar q_L\varepsilon,\bar\ell_L(\,)\bar\ell_L\varepsilon)\}$ where we insert $\psi_R^I$ in $(\,)$ and $SU(2)_L$ indices are summed with the antisymmetric 2-tensor. Finally we note that the operators with a superscript $*$ are genuinely CP odd~\footnote{One has, under CP, $h\to h, \varphi\to -\varphi \Rightarrow U\to U^*$, see \cite{Gavela:2014vra}.}.

The RGE for the coefficients of these operators is, for the set of bosonic operators,
\begin{align}
\mu\frac{d C_{1}}{d\mu}&=-\frac32\left(FF^{\prime\prime}\right)^2,  &
\mu\frac{d C_{2}}{d\mu}&=-\frac23\left(F^{\prime2}-1\right)^2, \end{align}
\begin{align}
\mu\frac{d C_{3}}{d\mu}&=-\frac12\left(FF^{\prime\prime}\right)^2-\frac13\left(F^{\prime2}-1\right)^2,\\
\mu\frac{d C_{4}}{d\mu}&=\frac{(FF^{\prime\prime})^2}{3}-2(F^{\prime2}-1)FF^{\prime\prime},
\end{align}\begin{align}
\mu\frac{d C_{5}}{d\mu}&=-\frac43 \left(FF^{\prime\prime}\right)^2, & 
\mu\frac{d C_{6}}{d\mu}&=\frac13\left(F^{\prime2}-1\right),\\
\mu\frac{d C_{7}}{d\mu}&=\frac23F^\prime FF^{\prime\prime} ,&
\mu\frac{d C_{8}}{d\mu}&=\frac16\left(F^{\prime2}-1\right),
\end{align}
whereas boson-Yukawa operators read,
\begin{align}
\mu\frac{d C_{9}^I}{d\mu}&=3 F^{\prime\prime}\left(F^\prime \mathcal{Y}^\prime_I -\frac{\mathcal{Y}_I}{F}\right), \\
 \mu\frac{d C_{10}^I}{d\mu}&=FF^{\prime\prime} \mathcal{Y}^{\prime\prime}_I+2\frac{F^{\prime2}-1}{F}\left(F^\prime \mathcal{Y}^\prime_I -\frac{\mathcal{Y}_I}{F}\right),\\\nonumber
  \mu\frac{d C_{11}^I}{d\mu}&= -2FF^{\prime\prime}\left(\frac {\mathcal{Y}_I} {F}\right)^\prime,
\end{align}
which should be read as equations for matrices $C$ in flavor space $dC_{\alpha\beta}\propto (\mathcal Y_I)_{\alpha\beta}$.
The scalar current times fermion current coefficients RGE is
\begin{align}
\mu \frac{dC_{12}^I}{d\mu}&=-\frac {iF}{2}\Big[ \mathcal{Y}^{\prime\dagger}_I\mathcal{Y}^{\prime\prime}_I+3\frac{\mathcal{Y}^\dagger_I}{F}\left(\frac {\mathcal{Y}_I}{F}\right)^\prime\Big] +h.c.,\\ \nonumber
\mu \frac{d C_{13}^I}{d\mu}&=-\frac {iF}{2}\Big[ \mathcal{Y}^{\prime}_I\mathcal{Y}^{\prime\prime\dagger}_I
+{\frac{\mathcal{Y}_I}{F}\left(\frac {\mathcal{Y}^\dagger_I}{F}\right)^\prime+ 2\frac{\tilde {\mathcal{Y}}_I}{F}\left(\frac {\tilde {\mathcal{Y}}^\dagger_I}{F}\right)^\prime}
\Big]+h.c.,\\ \nonumber
 \end{align}
 \begin{align}
\mu\frac{ dC_{14}^{IJ}}{d\mu}&=\frac{F^\prime}{F}\left(\mathcal{Y}^\dagger_I \mathcal{Y}_J\right)^\prime-\mathcal{Y}^{\prime\dagger}_I \mathcal{Y}^\prime_J-\frac{\mathcal{Y}^\dagger_I \mathcal{Y}_J}{F^2},\\ \nonumber
\mu \frac{dC_{15}^I}{d\mu}&=\left(\frac{F^\prime}{F}\left(\mathcal{Y}_I \mathcal{Y}^\dagger_I \right)^\prime- \mathcal{Y}^{\prime}_I\mathcal{Y}^{\prime\dagger}_I-\frac{\mathcal{Y}_I \mathcal{Y}^\dagger_I}{F^2}
\right),
\\ \nonumber
\mu \frac{dC_{16}^I}{d\mu}&=-\frac i2\left(\frac{F^\prime}{F}\mathcal{Y}_I\mathcal{Y}_I^{\prime\dagger} -\frac{F^\prime}{F}(\tilde {\mathcal{Y}}\tilde {\mathcal{Y}}^{\prime\dagger})_I\right)+h.c.,
\end{align}
 where again $C$ are matrices in flavor space so $\mathcal Y\mathcal Y^\dagger$ is the matrix product
and one can see that the CP odd operators are generated if the Yukawas $\mathcal Y$ have an imaginary component.
Lastly the four-fermion coefficients running reads,
 \begin{align} 
\mu\frac{d C_{17}^{IJ}}{d\mu}=&\, 2\left(\frac{\mathcal{Y}_I}{F}\right)^{\prime}\!\!\otimes\!\left(\frac{\mathcal{Y}^\dagger_J}{F}\right)^{\prime}-\mathcal{Y}^{\prime\prime}_I\otimes\mathcal{Y}^{\prime\prime\dagger }_J,\\ \nonumber
  &-\frac{3}{F^2}\left(F^\prime \mathcal{Y}^\prime_I -\frac{\mathcal{Y}_I}{F}\right)\!\otimes\!\left(F^\prime \mathcal{Y}^{\dagger\prime}_J -\frac{\mathcal{Y}^\dagger_J}{F}\right),
\\
 \mu\frac{d C_{18}^{IJ}}{d\mu}=& \left(\frac{\mathcal{Y}_I}{F}\right)^{\prime}\!\!\otimes\!\left(\frac{\mathcal{Y}_J}{F}\right)^{\prime} -\frac12\mathcal{Y}^{\prime\prime}_I\otimes\mathcal{Y}^{\prime\prime}_J,\\ \nonumber
&-\frac{3}{2F^2}\left(F^\prime \mathcal{Y}^\prime_I -\frac{\mathcal{Y}_I}{F}\right)\!\otimes\! \left(F^\prime \mathcal{Y}_J^\prime -\frac{\mathcal{Y}_J}{F}\right),
\end{align}
\begin{align} \nonumber
\mu\frac{d C_{19}^{IJ}}{d\mu}&=-4\left(\frac{\mathcal{Y}^\dagger_I}{F}\right)^{\prime}\!\!\otimes\!\left(\frac{\mathcal{Y}_J}{F}\right)^{\prime}\!, \\
\mu\frac{d C_{20}^{IJ}}{d\mu}&=-2\left(\frac{\mathcal{Y}_I}{F}\right)^{\prime}\!\!\otimes\!\left(\frac{\mathcal{Y}_J}{F}\right)^{\prime}\!,
\end{align}
where by tensor product we mean explicitly $d C_{\alpha\beta\gamma\delta}\propto \mathcal Y_{\alpha\beta}\otimes \mathcal Y_{\gamma\delta}$ noting that the order matters.
\section{Conclusions}
When the SM is extended to a general scalar geometry in HEFT, renormalizability --in the old acceptation-- is badly lost. In this work we made quantitative this feature providing the RGE induced by the leading --custodially preserving-- Lagrangian; from the point of view of HEFT, the SM limit is a very special one in which the the number of RGE equations collapses to a small linearly independent set. This feature can be used experimentally:  examining how the running of different couplings correlates is a tool in our quest to uncover the dynamics behind the scalar boson found at LHC.

The computation was carried out with a CDE and functional methods and along the way we derived the renormalization for an arbitrary manifold of scalar and Dirac fermions coupled via Yukawa-like interactions and an arbitrary gauge group.

\section*{Note added in proof}

In the process of readying this work for publication, and after the authors talked about the project at thesis defenses~\cite{ThesisKirill,ThesisSara} and more recently a dedicated seminar~\cite{SeminarUZH}, the paper~\cite{Buchalla:2017jlu} appeared on the same subject and with a large overlap. In~\cite{Buchalla:2017jlu} the one-loop UV divergent terms in HEFT are presented yet the explicit renormalization and applications are left for a companion paper; here instead we provide the RGE, include explicit examples of phenomenological relevance and our general formulae are valid for an arbitrary scalar manifold. 

\acknowledgments
S.S. acknowledges partial financial support by the European Union through the FP7 ITN
INVISIBLES (PITN-GA-2011-289442), by the Horizon2020 RISE InvisiblesPlus
690575, by CiCYT through the project FPA2012-31880, and by the Spanish
MINECO through the Centro de excelencia Severo Ochoa Program under grant SEV-2012-
0249 (GMS). The work of K.K. was supported by an ESR contract of the European Union network FP7 ITN INVISIBLES
(Marie Curie Actions, PITN-GA-2011-289442), of MICINN, through the project
FPA2012-31880 and by the University of Padova.
The work of S.S. was supported through the grant BES-2013-066480 of the Spanish MICINN.

\appendix\section{Reference formulae\label{appA}}

The Goldstone-dependent quantities $V_\mu$ and $T$ are often found in the literature, the connection with our notation is:
\begin{align}
&V_\mu \equiv (D_\mu U) U^\dagger =\frac{2i}{g}\sigma_A d_\mu \varphi^a t_a^A,\\
&T\equiv U\sigma_3U^\dagger=\frac{4}{gg_{\scriptscriptstyle Y}}t_Y^a t_a^A \sigma_A\,.
\end{align}
In the unitary gauge ($U.G.$) the metric is flat $g_{ab}=\delta_{ab}/v^2$ and
\begin{align}
&d_\mu\varphi^a_{U.G.}=\frac v2\left(gW^1_\mu\,,\,gW^2_\mu\,,\,gZ_\mu/c_\theta\right),\\
&[t_A^a]_{U.G.}=\frac{gv}{2}\delta_A^a,\qquad[t_Y^a]_{U.G.}=-\frac{g_{\scriptscriptstyle Y}v}{2}\delta_3^a.
\end{align}
where $\tan\theta=g_{\scriptscriptstyle Y}/g$ as customary.


\bibliography{LibRgeGYM}

\end{document}